\documentclass[proceedings]{rmaa}

\usepackage{paralist}

\title{Line-of-sight statistical methods for turbulent medium: VCS for emission
and absorption lines}
\author{
D. Pogosyan, 
 \altaffil{University of Alberta, Edmonton, Canada, pogosyan@ualberta.ca}
A. Lazarian  
 \altaffil{University of Wisconsin, Madison, USA, lazarian@astro.wisc.edu}
}

\listofauthors{D.~Pogosyan, A.~Lazarian}
\indexauthor{Pogosyan, D.}
\indexauthor{Lazarian, A.}

\resumen{}
\abstract{
We present an overview of the Velocity Coordinate Spectrum (VCS),
a new technique for studying astrophysical turbulence
that utilizes the line-of-sight statistics of Doppler-broadened spectral lines.
We consider the retrieval of turbulence spectra from
emission intensity observations of both high
and low spatial resolution and find that the VCS allows one
to study turbulence even when the emitting turbulent volume is not spatially
resolved. This opens interesting prospects for using the technique for 
extragalactic research. 
VCS developed for spectral emission lines is applicable to absorption lines
as well if the optical depth is used instead of intensity.
VCS for absorption lines 
in point-source spectra benefit from effectively narrow beam and does
not require dense sky coverage by sampling directions.
Even strongly saturated absorption lines still carry the information
about the small scale turbulence, albeit limited to the wings of a line.
Combining different absorption lines one can develop tomography of
the turbulence in the interstellar gas in all its complexity.

}


\shorttitle{Velocity Coordinate Spectrum for turbulent medium }
\shortauthor{Pogosyan \& Lazarian}

\begin{document}
\maketitle

\section{Introduction}

Turbulence is a key element of the dynamics of astrophysical
fluids, including those of interstellar medium (ISM), clusters of galaxies and 
circumstellar regions. The realization of the importance of turbulence
induces sweeping changes, for instance, in the paradigm of ISM. 
It became clear, for instance, that turbulence affects
substantially star formation, mixing of gas, transfer of heat.
Observationally it is known that the ISM is turbulent 
on scales ranging from AUs to kpc \citep[see][]{2004ARA&A..42..211E},
with an embedded magnetic field that influences almost all of its properties.

Using a statistical description is a nearly indispensable strategy when
dealing with turbulence.
One of the most widely used statistical measures is the turbulence
spectrum, which describes the distribution of turbulent fluctuations over
scales. For instance, the famous Kolmogorov model of incompressible turbulence
 predicts that the difference in velocities at
different points in turbulent fluid increases on average
with the separation between points as a cubic root of the separation,
i.e. $|\delta v| \sim l^{1/3}$. In terms of direction-averaged
energy spectrum this gives the famous Kolmogorov
scaling $E(k)\sim 4\pi k^2 P({\bf k})\sim k^{5/3}$, where $P({\bf k})$ 
is a {\it 3D} energy spectrum defined as the Fourier transform of the
correlation function of velocity fluctuations $\xi_v ({\bf r})=\langle  
\delta v({\bf x})\delta v({\bf x}+{\bf r})\rangle$.
Other types of turbulence, i.e. the turbulence of non-linear waves
or  the turbulence of shocks, are characterized by different power
laws and therefore can be distinguished from the Kolmogorov turbulence
of incompressible eddies.
Substantial advances in our understanding of the scaling of compressible
MHD turbulence (see review by \citealp{ChoLazarian2005},
and references therein) allows us to provide a direct comparison of
theoretical expectations with observations.

Recovering the velocity spectra from observations is a challenging problem.
Observations provide integrals of either emissivities or
opacities, both proportional to the local densities, 
in Position-Position Velocity space (henceforth PPV), i.e as a 
function of the direction on the sky and frequency that can be viewed
as the position in velocity coordinate along the line of sight.
Statistical measurements in this space are affected both by inhomogeneous 
spatial distribution and the motion of the matter. 
Turbulence is associated with fluctuating velocities that cause
fluctuations in the Doppler shifts of emission and absorption
lines.
If turbulence is supersonic, the recovering of its velocity spectrum is
possible with the Velocity Channel Analysis (VCA) 
of the emission intensity maps
\citep[LP00,LP04]{2000ApJ...537..720L,2004ApJ...616..943L}.
This technique has been successfully applied in several papers starting
with \citet{2001ApJ...551L..53S}.
However, the VCA is just one way to use
the general description of fluctuations in the PPV space. 
One can also study fluctuations along the {\it velocity coordinate}.  
The corresponding technique was termed the Velocity Coordinate Spectrum 
in \citet{2004JKAS...37..563L} and has been developed in 
\citet[LP06]{2006ApJ...652.1348L} for emission and
in \citet[LP08]{2008ApJ...686..350L}
for absorption lines. In this presentation we give an overview
of these developments.

\section{Linking observational and turbulent statistical measures}

Let us consider a volume of turbulent gas or
plasma (a ``cloud'') which extent along the line of sight $S$ is much smaller
than the distance to the observer.  Along a line of sight that
is labeled by a two-dimensional position vector ${\bf X}$ on the cloud
image, the intensity in a spectral (either emission or absorption) 
line with rest-frame frequency $\nu_0$, $I_{\bf X}(\nu,\nu_0)$, is
determined by the density of emitting (or absorbing) atoms in PPV space
(LP06,LP08),
$\rho_s({\bf X},v)$
\begin{eqnarray}
I_{\bf X}(v) &=&
\frac{\epsilon}{\alpha}\left[1-{\mathrm e}^{-\alpha \rho_s({\bf X},v)}\right] 
 \quad emission
\label{simplified} \\
I_{\bf X}(v) &=& I_0 {\mathrm e}^{-\alpha \rho_s({\bf X},v)} 
\quad\quad\quad absorption
\label{simplified:abs}
\end{eqnarray}
where $\epsilon$ is the emissivity and $\alpha$ is the absorption 
\footnote{Self-absorption in case of emission line} coefficient.
and the velocity $v \approx \frac{c}{\nu_0} (\nu-\nu_0)$, if the intrinsic line
width is neglected.

In the simplest case of optically thin emission lines and unsaturated 
absorption lines the PPV density is directly connected to observables,
the emission intensity or the optical depth for absorption
\begin{eqnarray}
\rho_s({\bf X},v) &\propto & I_{\bf X}(v), 
\quad\quad \alpha \to 0, \quad emission  \\
\rho_s({\bf X},v) &\propto& \tau_{\bf X}(v) \equiv -\ln(I/I_0), 
~ absorption
\end{eqnarray}

PPV density $\rho_s({\bf X},v)$ encodes the information
about both the turbulent velocities and the inhomogeneity in
the matter distribution. The characteristic feature that allows us to 
distinguish between the thermal and turbulent velocity is the spatially 
correlated nature of the latter.
Under set of assumptions employed in LP00 and LP04,  
the PPV structure function
$d_s(R,v)=\langle \left[\rho_s({\bf X_1},v_1)-\rho_s({\bf X_2},v_2)\right]^2\rangle$,
$R=|{\bf X_1}-{\bf X_1}|,~ v=|v_1-v_2| $,
can be expressed via real space velocity
and density correlations as
\begin{equation}
d_s(R,v) \propto
\int_{-S}^S \!\!\!\!\! {\mathrm d}z 
\; \frac{\xi( r)}{\sqrt{D_z({\bf r})+2\beta}}
\exp\left[-\frac{v^2}{2 ( D_z({\bf r}) + 2 \beta) }\right],
\label{ksicloud}
\end{equation}
where the structure function of the velocity $z$-components is  
\begin{equation}
D_z({\bf r}) \approx Cr^{m}~~,
\end{equation}
( $C$ is a normalization constant ) and for 
the density correlation function 
$\xi({\bf r}=|{\bf x_1} - {\bf x_2}|)=\langle \rho({\bf x_1})
\rho({\bf x_2})\rangle$
the isotropic power-law form is adopted 
\begin{equation}
\xi(r) \propto
\left(1 + \left[ {r_0 \over r} \right]^\gamma\right) ~~ .
\label{eq:xi}
\end{equation}
The thermal broadening is described by $\beta=\frac{k_{\mathrm B} T}{m_a}$.

The Eq.~(\ref{ksicloud}) provides way for recovery of the turbulence
by measuring the correlations of the observed signal, for example the
structure functions
\begin{equation}
\left.
\begin{array}{l}
{\cal D}(R,v) \equiv 
\langle \left[ I_{\bf X_1}(v_1) - I_{\bf X_2}(v_2) \right]^2 \rangle  \\
{\cal D}(R,v) \equiv 
\langle \left[ \tau_{\bf X_1}(v_1) - \tau_{\bf X_2}(v_2) \right]^2 \rangle
\end{array}
\right\}
\propto d_s(R,v)
\label{eq:DRv}
\end{equation}

Despite formal similarity, use of emission and absorption
lines has also several distinct features.

Measuring the emission by the turbulent gas typically provides us with
a sky map of the signal. If the angular resolution of such map is high,
it is more advantageous
to measure angular correlations across the image ${\cal D}(R,0)$.
This leads to the VCA analysis (LP00, LP04) that has been reviewed 
at the ``Magnetic Fields in the Universe I'' meeting
\citep{2005AIPC..784..287P}. The use of the line-of-sight measure
${\cal D}(0,v)$ that we review here is more challenging  but has an advantage
for the data with poor angular resolution.

Absorption spectra, on the other hand, typically represents real pencil beams
for which the line-of-sight statistics is the only one available. The 
effective angular resolution of such spectra is high if the source is 
a point object. Therefore both high and low angular resolution limits
of the VCS are of interest. One can also applying VCS to the wings of 
saturated absorption lines (LP08) although the issues of noise, 
velocity resolution and intrinsic line broadening become critical
for such studies.

\section{VCS for a Transparent Emitting Medium}
The intensity in optically thin lines provides direct information
on the density in PPV space according to Eq.~(\ref{eq:DRv}).
Finite angular resolution of the instrument leads to the beam-averaged
measurements
\begin{equation}
{\cal D}(0,v) = \int d {\bf R} B \left({\bf R} \right) d_s (R,v) ~.
\end{equation}
where the realistic beam has a finite width, $\Delta B$. 

We shell formulate VCS results using the velocity coordinate power spectrum
\begin{equation}
P(k_v) = \int d v {\cal D}(0,v) \exp\left[-k_v^2 v^2/2 \right]
\end{equation}
where $k_v$ is the ``wave-number'' reciprocal to the velocity $v$ coordinate.
With regard to the observation scale $k_v$, two other scales define
the different regimes of the analysis:

From instrumental perspective it is resolution scale is important.
To the linear scale $\Delta B$ corresponds the velocity scale 
\begin{equation}
V_{\Delta B} \equiv \sqrt{D_z(S) (\Delta B/S)^m}~~,
\end{equation}
equal to the magnitude of turbulent velocities at
the separation of a size $\Delta B$. For a measurements at the scale $k_v$
the beam is effectively narrow when
\begin{equation}
k_v < { V_{\Delta B} } ^{-1}
\label{eq:nw}
\end{equation}
while on shorter scales its width is important. Note that it is angular
resolution that affects the behaviour of line-of-sight statistics.
Similarly, for 2D VCA analysis in channel maps velocity resolution played a
role in interpreting 2D statistics of intensity maps

From the perspective of the turbulence properties
the amplitude of the density
contribution to VCS is encoded in the correlation length $r_0$.
This scale decides whether density inhomogeneities dominate in Eq.~(\ref{eq:xi})
or the turbulence effects are manifest only in random motions of the uniform
gas distribution, $\xi(r)=1$.
The velocity scale that corresponds to $r_0$ is
\begin{equation}
V_{r_0} = \sqrt{D_z(S) \left(r_0/S\right)^m}~~~,
\end{equation}
So we can roughly equate the scale of equality of the density and velocity
effects to $r_0$, i.e., in velocity units, 
\begin{equation}
k_v \approx {V_{r_0}}^{-1} ~~.
\end{equation}

\begin{figure*}[!t]
 \includegraphics[width=0.7\columnwidth]{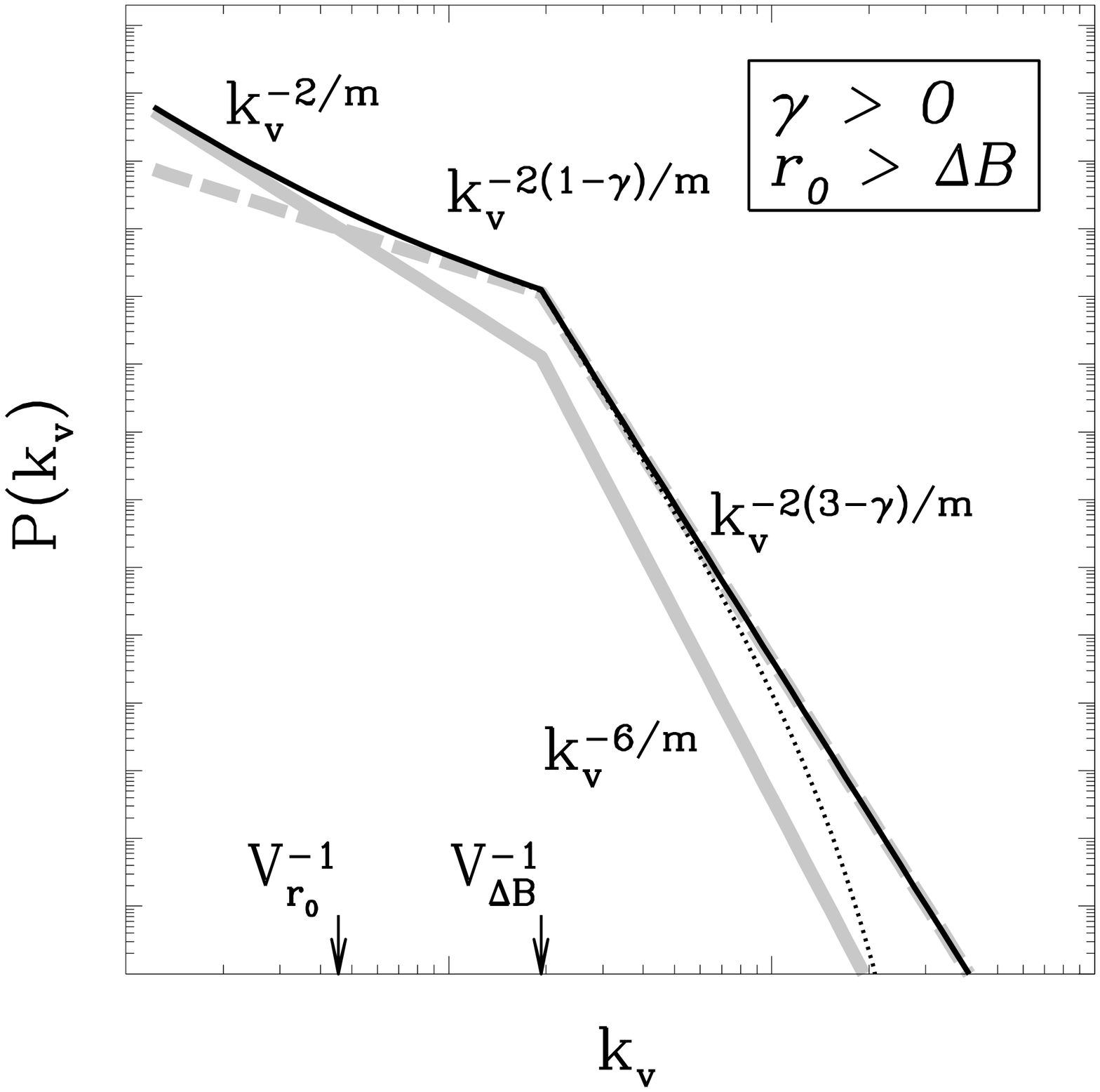}%
 \hspace*{0.3\columnsep}%
 \includegraphics[width=0.7\columnwidth]{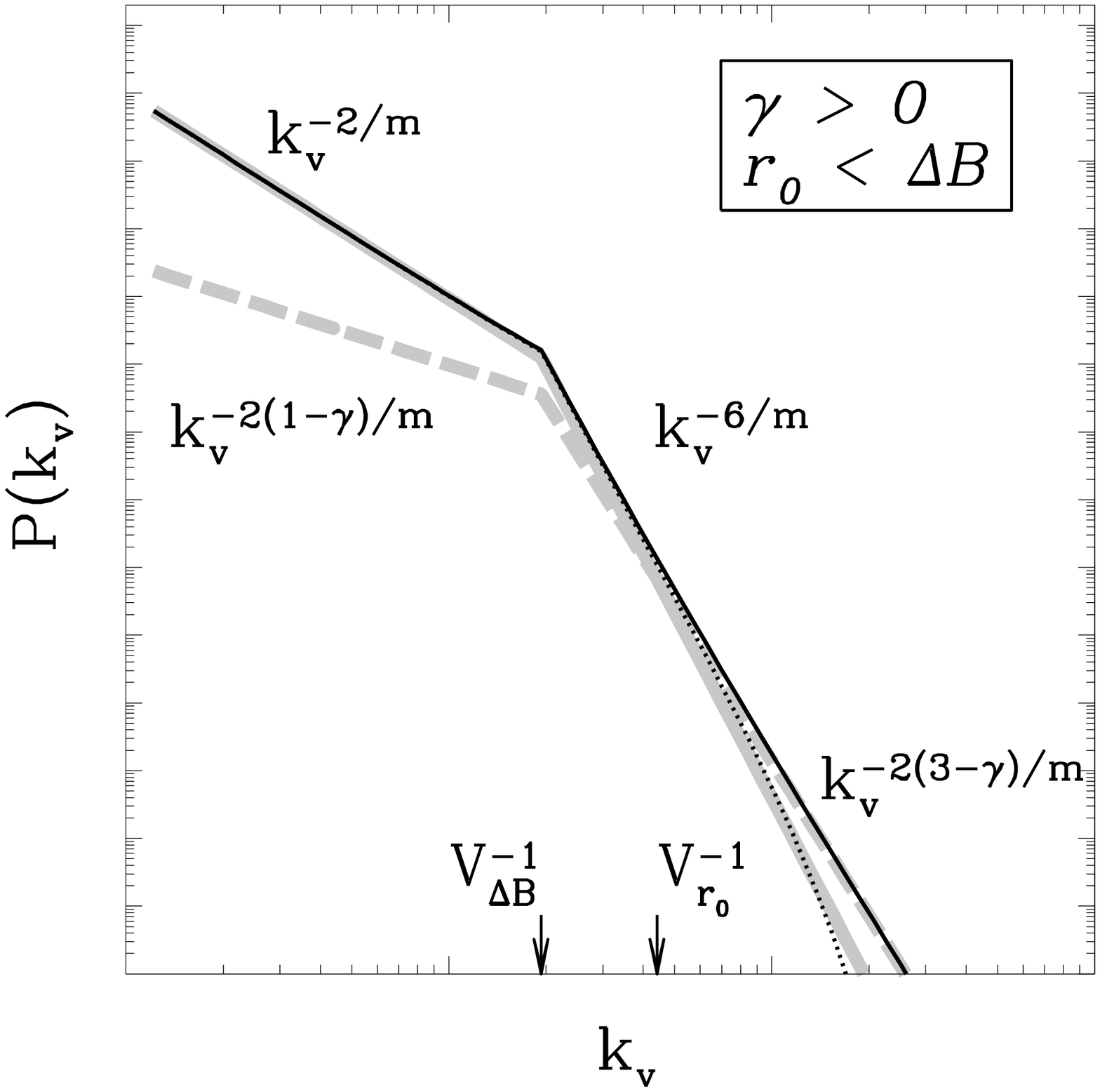}%
 \hspace*{0.3\columnsep}%
 \includegraphics[width=0.7\columnwidth]{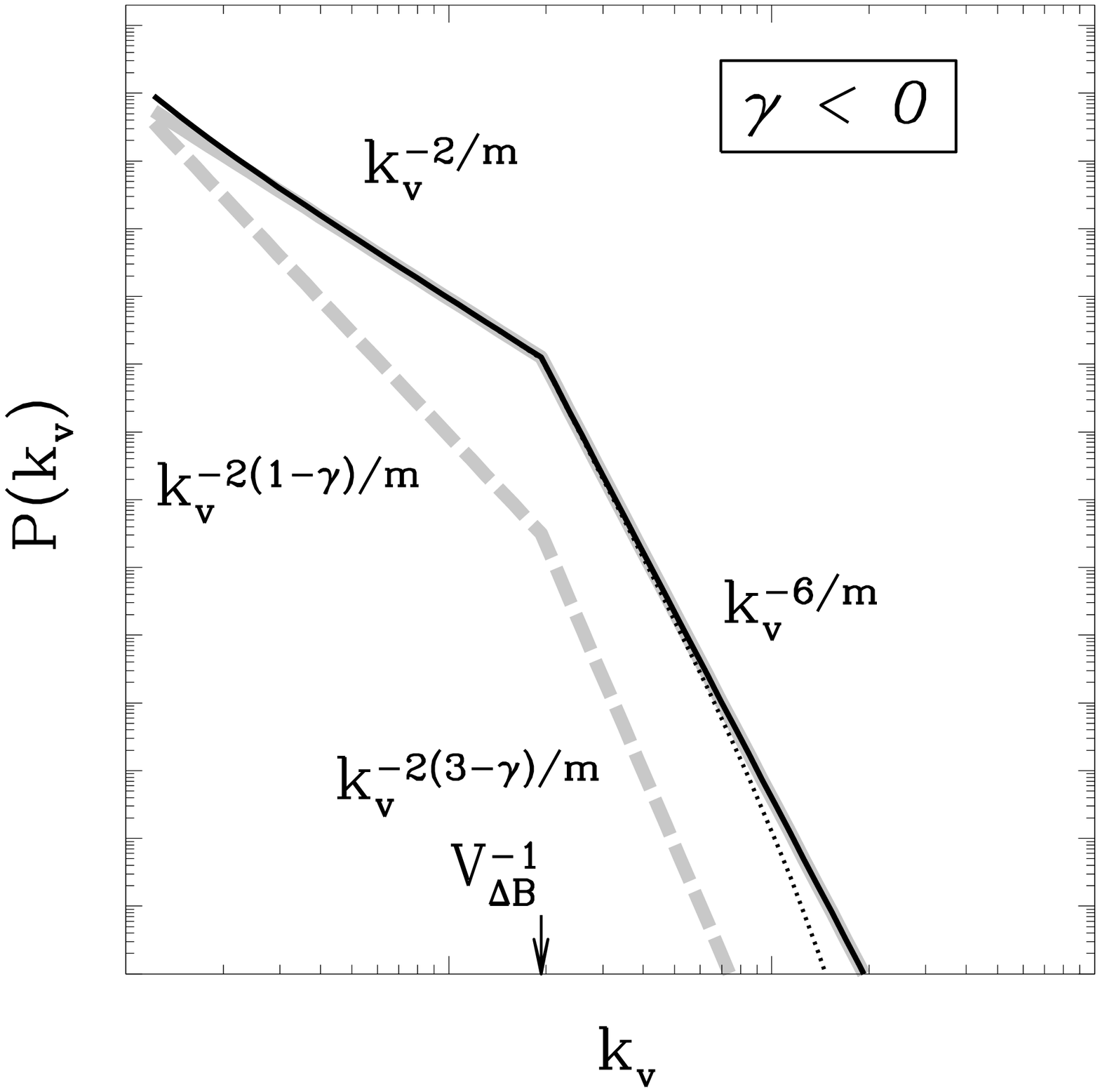}%
\caption{ Qualitative representation of the density and velocity contributions
to the VCS power spectrum and the resulting scaling regimes.
In every panel light lines show contributions from the $\rho$-term
(density modified by velocity, dashed line) and $v$-term (pure velocity
effect, solid line) separately, while the dark solid line shows the combined
total VCS power spectrum. Thermal suppression of fluctuations
is shown by the dotted line. The labels above the dark solid curve
are arranged so as to illustrate
the sequential transition of the scalings of the total power spectrum.
Everywhere except the intermediate regimes, the total spectrum is dominated by
one of the components to which the current labeled scaling corresponds.
Labels below the dark solid lines mark the scaling of the subdominant
contributions.
For the left and middle panels the density
power spectrum is taken to be shallow, $\gamma > 0$.
The left panel corresponds to high amplitude of the density correlations,
$r_0 > \Delta B$, where density effects become dominant at relatively
long wavelengths for which the beam is narrow. In the middle panel,
the amplitude of density correlations is low $r_0 < \Delta B$ and they
dominate only the smallest scales which results in the intermediate steepening
of the VCS scaling. The right panel corresponds to the  steep density spectrum.
In this case the density contribution is always subdominant.
In this example the thermal scale is five times shorter than
the resolution scale $V_{\Delta B}$.
}
\label{fig:1}
\end{figure*}

In Figure~\ref{fig:1} we summarize the different scalings of VCS.
As for the VCA the main difference stems from
the density being either shallow or steep. If the density is shallow
i.e. scales as $\xi\sim r^{-\gamma}$, $\gamma>0$, which means
that the correlations increase with the decrease of the scale,
then it eventually becomes important at sufficiently small velocity
differences, i.e. at sufficiently large $k_v$. In the opposite case, i.e. when
$\gamma<0$, the contributions of density can be important only at
large velocity separations. 

The left and middle panels of Figure~\ref{fig:1} deal with the case of shallow density. 
Velocity is dominant at $k_v < k_0$, while the density term provides the 
main contribution at $k_v > k_0$. The left panel demonstrates the case where
the scale of transition from asymptotic is entirely
dominated by velocity to the one influenced by spatially resolved velocity and density, $V_{\Delta B} < V_{r_0}$.
The observed fluctuations arising from
 the unresolved turbulent eddies depends on the scalings of both
velocity and density. In the middle panel, $ V_{\Delta B} > V_{r_0}$,
the transition scale is unresolved. In this case
if there is still a dynamical range for
moderately long scales to be resolved by the experiment
$D_z(S)^{1/2} > k_v^{-1} > V_{\Delta B}$, 
the VCS of the resolved eddies will be determined by the turbulent velocities
only.

The right panel of Figure~\ref{fig:1} addresses the case of a steep density
spectrum. The difference now is that fluctuations of the density are maximal
at low wave-numbers and it is there that the density
could be important.  Velocity is dominant at shorter scales $k_v > k_0$.
However,  the steep density correlation length $r_0$ is large,
at least as large as the density power cutoff $r_c$,
which argues for density fluctuations to be
subdominant everywhere up to the scale of the emitting turbulent volume
(``cloud'').

\subsection{Thermal effects and inertial range}

In Figure~\ref{fig:1} we have plotted theoretical
power spectra over 3 decades of
velocity magnitude to compactly demonstrate all possible scalings.
Thermal effects are shown as well, which in a language of the power spectrum
are simply described by $\exp[-\beta k_v^2]$ cutoff factor.

In observations, one can  anticipate a coverage over
two decades of velocity magnitudes before thermal effects
get important. Potentially, correcting for the thermal
broadening,
one can extend the observational range further. Note, that the thermal
corrections are different for species of different mass. Therefore, by using
heavier species one can extend the high $k$ cut-off by the square root of
the ratio of the mass of the species to the mass of hydrogen. 

How deeply subsonic scales can still be probed in cold gas depends
on the signal-to-noise ratio of the available data. Indeed, 
the ability to deconvolve the thermal smoothing is limited by noise
amplification in the process.
However, any extension of the VCS by a factor $a$  in velocity
results in the extension of the sampled spatial scales by a factor of
$a^{2/m}$, which is $a^3$ for the Kolmogorov turbulence.  
   
Even with a limited $k_v$ coverage, important results can be obtained
with VCS, especially if observations encompass one of
the transitional regimes.
For example, if one measures a transition from
a shallow spectrum to a steeper one (see the left and the right panels
in Figure~\ref{fig:1})
one has the potential to i) determine the velocity index $m$,
since the difference between the slopes is always $4/m$;
ii) determine $\gamma$ next;
iii) estimate the amplitude of turbulent velocities from the position
of the transition point as discussed above.
On the other hand, if one encounters a transition from a steep to a shallower
spectrum, one i) may argue for the presence of the shallow density
inhomogeneities;
ii) estimate $\gamma/m$ from the difference of the slopes, and then $m$;
iii) estimate the density correlation radius $r_0$. Finally, if no transition
regime is available, 
then for $\gamma <0$ one can get $m$, while for the case
of $\gamma>0$ a combination of $\gamma$ and $m$ is available.

\section{VCS for Absorption lines}
Studies of turbulence with absorption lines are possible with the VCS technique
if, instead of intensity $I(\nu)$, one uses the optical depth $\tau (\nu)$,
i.e., the logarithm of the absorbed
intensity $\log I_{abs}(\nu)$.

In the weak absorption regime, i.e. when the optical depth at the middle of the absorption line is less than unity, the analysis
of the  $\tau(\nu)$ coincides with the analysis of emission discussed above,
taken in the limit of ideal angular resolution.

\begin{figure}[!t]
\centerline{\includegraphics[height=.75\columnwidth]{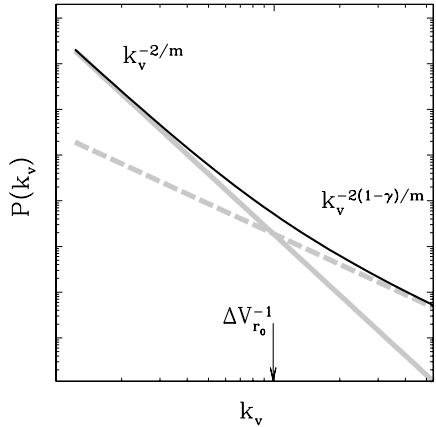}
}
\caption{Power spectrum of the optical depth fluctuations due to
turbulent motions of absorbers in unsaturated absorption 
line. Thermal and intrinsic line smoothing is factored out. 
Effect of density inhomogeneities becomes dominant at $k_v > V_{r_0}^{-1}$,
assuming shallow density, $\gamma > 0$. 
}
\label{fig:Ptot}
\end{figure}

In Figure~\ref{fig:Ptot} we summarize different regimes for
studying turbulence using absorption lines from point sources for which angular
resolution can be considered as perfect. For shallow density, i.e. for 
$\xi(r) \approx \bar n^2 \left[1+ (r_0/r)^{\gamma}\right]$, $\gamma>0$
the spectrum of optical depth
fluctuations that  arise from inhomogeneities in density projected onto
velocity coordinate scales as $P_\rho(k_v)\sim k_v^{-2(3-\gamma)/m}$. 
For sufficiently large $k_v$ its contribution dominates the
contribution $P_v(k_v)\sim k^{-2/m}$ coming from purely velocity 
projection effect of 
turbulently moving uniformly distributed, $\xi(z)\approx \bar n^2$, gas.
If we measure the point $\Delta V_{r_0} \approx \sqrt{D_z(S)}(r_0/S)^{m/2}$ 
at which $P_\rho$ becomes dominant over $P_v$,
we can estimate the density
correlation radius $r_0$ which has the physical meaning of the scale at which
the dispersion of the fluctuations of density equal the mean density
(see LP06). 
However, whether we can observe both regimes or just one, depends on the mask
that is imposed by the saturation of the absorption line, as well as the
small scale filtering arising from thermal broadening. 
For $\gamma<0$, $P_v$ always dominates.

In the intermediate absorption regime, i.e. when the optical depth at the
middle of the absorption line is  larger than unity, but less than $10^3$,
the wings of the absorption line can be used for the analysis. The central,
saturated part of the line is expected to be noise dominated.

Higher the absorption, less the extend of the wings available for the analysis
is. In terms of
the mathematical setting this introduces and additional window in the expressions for the VCS analysis. However, the contrast
of the small scale fluctuations increases  with the decrease of the window.

For strong absorption regime, the broadening is determined by Lorentzian wings of the line and therefore no  information
on turbulence is available.

\section{VCS cookbook}

The VCS cookbook  for emission lines is rather straightforward. VCS
near a scale $k_v$ depends on whether the instrument  
resolves the correspondent spatial scale
$\left[k_v^2 D_z(S)\right]^{-1/m} S$,
where $S$ is the scale where turbulence saturates. 
If this scale is resolved then $P_v(k_v) \propto k_v^{-2/m}$
and $P_{\rho}(k_v) \propto k_v^{-2(1-\gamma)/m}$. 
If the scale is not resolved then
$P_v(k_v) \propto k_v^{-6/m}$ and $P_\rho(k_v) \propto k_v^{-2(3-\gamma)/m}$. 
These results are presented in a compact form in Table~\ref{table:results}.
\begin{table*}
\centering
\setlength{\tabcolsep}{3.8\tabcolsep}
\caption{Scalings of VCS $P(k_v)=P_\rho(\lowercase{k_v})+P_v(\lowercase{k_v})$
for shallow and steep densities.}
\label{table:results}
\begin{tabular}{lll} \toprule \\[-2mm]
Spectral~term & $\Delta B < S \left[k_v^2 D_z(S)\right]^{-\frac{1}{m}}$ &
$ \Delta B > S \left[k_v^2 D_z(S)\right]^{-\frac{1}{m}}$ \\[2mm]
\midrule
$ P_\rho(k_v) $ & $ \propto\left(k_v D_z^{1/2}(S)\right)^{-2(1-\gamma)/m} $& 
$ \propto\left(k_v D_z^{1/2}(S)\right)^{-2(3-\gamma)/m} $ \\[2mm]
\midrule
$ P_v(k_v) $ & $ \propto\left(k_v D_z^{1/2}(S)\right)^{-2/m} $  & 
$ \propto\left(k_v D_z^{1/2}(S)\right)^{-6/m} $ \\[2mm]
\bottomrule
\end{tabular}
\end{table*}
The transition from the low to the high resolution regimes happens as
the velocity scale under study gets comparable to the turbulent velocity
at the minimal spatially resolved scale. As the change of slope is the
velocity-induced effect, it is not surprising that the difference in
spectral indexes in the low and high resolution limit is $4/m$ for both
$P_v$ and $P_\rho$ terms, i.e it does not depend on the density
This allows for separation of the velocity and density contributions.
For instance, Figure~\ref{fig:1}
illustrates that in the case of shallow density both the density and velocity 
spectra can be obtained.

Note that obtaining the density spectrum from a well resolved
map of intensities is trivial for the optically thin medium,
as the density spectrum is directly available from the column densities (i.e.
velocity integrated intensities). However, for the absorbing
medium such velocity-integrated 
maps provide the universal spectrum $K^{-3}$,
where $K$ is the 2D wavenumber (LP04). Similarly, even for the optically thin
medium, it is not possible to get the density spectrum if the turbulent
volume is not spatially resolved. On the contrary, $P_{\rho}(k_v)$ reflects
the contribution of shallow density even in this case.

An advantage characteristic to 
absorption lines is that they provide measurements in the
"high resolution regime", provided that the absorption against a point source
is used. This is important, in particular, for turbulence research
in extragalactic setting where even high angular resolution may not translate
into high spatial resolution.
To get proper statistical averaging, one requires more than one point source.
Numerical
experiments in \citet{2006astro.ph.11465C,2008arXiv0811.0845C} showed that
using measuring spectra along 5-10 lines of sight is enough to recover
the underlying spectrum properly.  

In the simplest case, Figure~\ref{fig:Ptot}
illustrates how the information on velocity
and density can be obtained from observations for the case of shallow density.
In the case of steep density the measured spectrum is always
$\sim k_v^{-2/m}$, which allows finding the spectral index of velocity $m$.

Similar to the case of emitting gas, the VCS study of absorbing gas is
dominated by the coldest gas component.
A practical emission data handling using the VCS technique in
\citet{2006astro.ph.11462C}
showed that fitting the observed power spectrum using the integral expressions
is a more reliable way of obtaining both parameters of the underlying
turbulence and temperature of the emitting gas. 

The difference in data analysis for absorption lines compared to that 
of \citet{2006astro.ph.11462C}
is twofold. Firstly, one has to take into account masking
of the data that restricts it to the wings region if the line is saturated.
Secondly, for emission maps analyzed in \citet{2006astro.ph.11462C} the
fitting was done while varying the spatial resolution of the data.
As the effects of particular factors, e.g. thermal broadening, are different
at different spatial scales, one can provide additional testing of the
accuracy of the fit by varying the angular resolution. Varying the resolution
is not an option for the absorption data from point sources but is
possible when spatially extended sources are used to sample
the medium. In the latter case the VCA-type studies of the absorption
lines wings may also be possible.

\end{document}